\documentclass{article}

\usepackage[preprint]{neurips_2023}
\usepackage[page]{appendix}

\usepackage[utf8]{inputenc} % allow utf-8 input
\usepackage[T1]{fontenc}    % use 8-bit T1 fonts
\usepackage{hyperref}       % hyperlinks
\usepackage{url}            % simple URL typesetting
\usepackage{booktabs}       % professional-quality tables
\usepackage{amsfonts}       % blackboard math symbols
\usepackage{nicefrac}       % compact symbols for 1/2, etc.
\usepackage{microtype}      % microtypography
\usepackage{xcolor}         % colors

\usepackage{xspace}
\usepackage{multirow}
\usepackage{tabularx}
\usepackage{amsmath}
\usepackage{mathtools}
\usepackage{colortbl}
\usepackage{tablefootnote}
\usepackage{soul}
\usepackage{pifont}
\usepackage{enumitem}

\usepackage{amsmath,amsfonts,bm}

\def\eqref#1{equation~\ref{#1}}
\def\1{\bm{1}}

\def\vh{{\bm{h}}}

\def\vk{{\bm{k}}}

\def\vo{{\bm{o}}}

\def\vq{{\bm{q}}}

\def\vs{{\bm{s}}}

\def\vz{{\bm{z}}}

\def\mA{{\bm{A}}}

\def\mW{{\bm{W}}}

\DeclareMathAlphabet{\mathsfit}{\encodingdefault}{\sfdefault}{m}{sl}
\SetMathAlphabet{\mathsfit}{bold}{\encodingdefault}{\sfdefault}{bx}{n}

\newcommand{\R}{\mathbb{R}}

\newcommand{\softmax}{\mathrm{softmax}}

\newcommand{\method}{MolEdit3D\xspace}

\title{Structure-Based Drug Design \\via 3D Molecular Generative Pre-training and Sampling}

\author{Yuwei ~Yang \\
  ByteDance Research\\
  \texttt{yuwei.yang@bytedance.com} \\
  \And
  Siqi ~Ouyang \\
  Language Technologies Institute \\ 
  Carnegie Mellon University \\
  \texttt{siqiouya@andrew.cmu.edu}
  \And
  Xueyu ~Hu \\
  Guanghua Law School \\
  Zhejiang University \\
  \texttt{huxueyu@zju.edu.cn}
  \And 
  Mingyue ~Zheng\\
  Shanghai Institute of Materia Medica \\ 
  Chinese Academy of Sciences \\
  \texttt{myzheng@simm.ac.cn}\\
  \And
  Hao ~Zhou \\
  Institute of AI Industrial Research \\
  Tsinghua University \\
  \texttt{zhouhao@air.tsinghua.edu.cn}\\
  \And 
  Lei ~Li \\ 
  Language Technologies Institute \\ 
  Carnegie Mellon University \\
  \texttt{leili@cs.cmu.edu} \\
}
 
\begin{document}

\maketitle

\begin{abstract}
    
Structure-based drug design aims at generating high affinity ligands with prior knowledge of 3D target structures. Existing methods either use conditional generative model to learn the distribution of 3D ligands given target binding sites, or iteratively modify molecules to optimize a structure-based activity estimator. 
The former is highly constrained by data quantity and quality, which leaves optimization-based approaches more promising in practical scenario. However, existing optimization-based approaches choose to edit molecules in 2D space, and use molecular docking to estimate the activity using docking predicted 3D target-ligand complexes. The misalignment between the action space and the objective hinders the performance of these models, especially for those employ deep learning for acceleration. In this work, we propose \method to combine 3D molecular generation with optimization frameworks. We develop a novel 3D graph editing model to generate molecules using fragments, and pre-train this model on abundant 3D ligands for learning target-independent properties. Then we employ a target-guided self-learning strategy to improve target-related properties using self-sampled molecules. \method achieves state-of-the-art performance on majority of the evaluation metrics, and demonstrate strong capability of capturing both target-dependent and -independent properties.

\end{abstract}

\section {Introduction}
    Drug molecules exhibit their activities by forming tightly binding 3D complex with disease-related targets. Rooted in this concept, structure-based drug discovery (SBDD) aims to design ligands (drug candidates) using the prior knowledge of the 3D target structure. \citep{batool2019structure} Ideally, identified ligand molecules should be 1) novel to existing database 2) satisfying target-independent properties, such as being easy-to-synthesize, drug-like, and energetically stable 3) satisfying target-dependent properties, particularly, demonstrating high binding affinity to the given target. 

It is challenging to design drugs satisfying the above criteria. 
Traditional SBDD employs  virtual screening to filter molecules from a large database \citep{bajorath2002integration,ferreira2015molecular, meng2011molecular}, which cannot identify new molecules. Recent work using deep generative models show promise in generating novel ligands. One widely used approach formulates the problem as conditional generation, which use 3D target-ligand complex data to learn the distribution of ligands given a target. These methods face two difficulties: 1) due to the nature of generative model, the generated ligands have similar structures and properties as those in the training data and may not surpass known actives,\citep{walters2020assessing} 2) experimental measured 3D target-ligand complexes are scarce, and may not be sufficient to support the learning of both target-dependent and -independent properties.\citep{liu2017forging}

Another line of work treats SBDD as an optimization task and employs binding affinity as an objective to reflect the fitness between ligands and targets. Molecular docking is commonly used for structure-based binding affinity estimation.\citep{meng2011molecular} It uses global optimization to identify the lowest binding energy (highest binding affinity) and its corresponding binding pose on the system's energy surface, which is called docking. If a 3D target-ligand complex is given, molecular docking can also directly calculate or conduct local minimization to estimate the binding affinity, which is called scoring or minimization respectively. AutoGrow models \citep{durrant2009autogrow,spiegel2020autogrow4},  use genetic algorithm as the optimization engine and iteratively modifies 2D ligands to achieve better docking score. RGA \citep{fu2022reinforced} uses reinforcement learning to improve the optimization efficiency of genetic algorithm and reduces its randomness in SBDD task. 
However, the above-mentioned methods generate 2D molecules and rely on docking, a computationally expensive oracle, to obtain the binding affinity, which greatly impact their efficiency. In addition, when using deep learning to accelerate the optimization procedure for SBDD, the misalignment between the objective and the action space can also influence the model performance. More specifically, the objective, namely docking score, reflects the 3D interactions between the target and the ligand, however, the editing operations are conducted on 2D molecules. 

To address these challenges in existing SBDD models, we propose a pre-trained 3D graph editing model (\method) to combine 3D molecular generation with target-guided optimization. As shown in Figure~\ref{fig:overview}, we designed a novel 3D graph editing model which generates 3D molecules by adding or deleting fragments. Using fragments as building blocks can improve the validity of local structures and reduce model complexity comparing to atom- and bond-based generation. 
The model is pre-trained using abundant 3D molecules to capture target-independent properties. We implement the 3D generation model in a Bayesian sampling procedure with simulated annealing to optimize desired target-dependent properties. In addition, we adopts self-learning to fine-tune the model using self-generated samples for the improvement of sampling efficiency and further enhancement of target-awareness. 

Our contributions are as follows: 
\begin{itemize}[itemsep=1pt, leftmargin=10pt, parsep=0pt, topsep=1pt]
    \item We develop a 3D molecular graph editing model to generate 3D molecules by modifying fragments.
    \item We develop techniques to pre-train the 3D graph editing model with 3D ligands and further fine-tune it using self-generated samples with target-guidance. 
    \item Our experimental results demonstrate that \method generates molecules with higher binding affinities compared to the best prior method (Vina score: -10.16 versus -9.77), and improves success rate by an absolute of 13.8\% from the previous best. In addition, the generated molecules maintain proper target-independent properties, including energetic stability and drug-like ring compositions. 
\end{itemize}

\section {Related Work}
    In this section, we review previous work focusing on conditional generation and optimization-based approaches for SBDD. Additional related work can be found in Appendix. %\ref{appendix: related_work}.

\textbf{Conditional 3D Drug Design} Generating 3D molecules is a relatively new research area in AI powered drug discovery due to its intrinsic difficulty. Within this field, conditional 3D drug design task requires the model to generate novel 3D molecules that are geometrically constrained by the target binding site and also satisfy multiple drug-like requirements. Comparing to generating 1D/2D molecules, the additional dimension significantly increases the explorable molecular space and makes this task more challenging. Regardless of its significance, limited efforts have been made in this field. \citet{masuda2020ligan} use variational autoencoders \citep{kingma2013auto} to learn the 3D ligand distribution conditioned on the target structures using \texttt{CrossDocked2020} \citep{francoeur2020three}, a 3D target-ligand complex database. Similarly, \citet{luo2021a} model the probability of atom occurrence within the target binding sites using mask prediction. \citet{pmlr-v162-liu22m} encode both protein and ligand, and places new atoms sequentially based on the contextual features while preserving the equivariance property. It should be noted, the above-mentioned methods can only model atoms and rely on a post-processing algorithm to assign bonds between the generated atoms, which is possible to create chemically invalid molecules. To improve validity, \citet{peng2022pocket2mol} use an equivariant generative network to predict both atoms and bonds conditioned on pocket features. \citet{long2022DESERT} propose to generate molecules conditioning on pocket shape instead of conditioning on the complete target structure. In addition, they use fragment as base unit to generate ligands and employ a greedy algorithm to connect the generated 3D fragments. Our method also generates 3D molecules.
It treats molecules as linked rigid fragments, which improves the chemical validity of local structures. In addition, it predicts the connectivity and the torsion angles between fragments which learns rational fragment connectivity and energetic stability from training data.

\textbf{Optimization-Based Drug Design} Drug design is essentially an optimization task, and previously deep models have used various approaches to optimize the properties, such as Bayesian optimization in latent space \citep{gomez2018automatic,jin2018junction, winter2019efficient}, reinforcement learning \citep{de2018molgan, popova2018deep, you2018graph, popova2019molecularrnn, Shi2020GraphAF, zhou2019optimization}, evolutionary and genetic algorithms \citep{ahn2020guiding, jensen2019graph, devi2015evolutionary,Nigam2020Augmenting,a2012multi}, sampling based approach \citep{fu2021mimosa, xie2021markov} and etc. The above-mentioned methods are developed for optimization in ligand-based drug design scenario, which neglects the 3D interactions between the ligands and the targets. \citet{spiegel2020autogrow4} and \citet{fu2022reinforced} adopt genetic algorithm for SBDD by incorporating molecular docking objective. Although the objective reflects 3D interactions, their action space is still defined on 2D graphs.

Molecular editing is an important component in optimization-based drug design. For instance, \citet{zhou2019optimization} edit 2D molecules by adding or deleting atoms and bonds, \citet{xie2021markov} and \citet{fu2021mimosa} edit 2D molecules by adding, deleting or replacing fragments, and \citet{Nigam2020Augmenting} edit 1D SELFIES strings \citep{krenn2020self} by inserting or replacing single atoms or phenyl rings. In this work, we further develop editing-based molecular generation for 3D drug design. More specifically, we propose to add and delete rigid fragments in 3D space and use torsion angles to describe the relative spatial location of the newly edited region.

\section {The Proposed Method}
    \begin{figure}[tb]
    \centering
    \vspace{-5mm}
    \includegraphics[width=\linewidth]{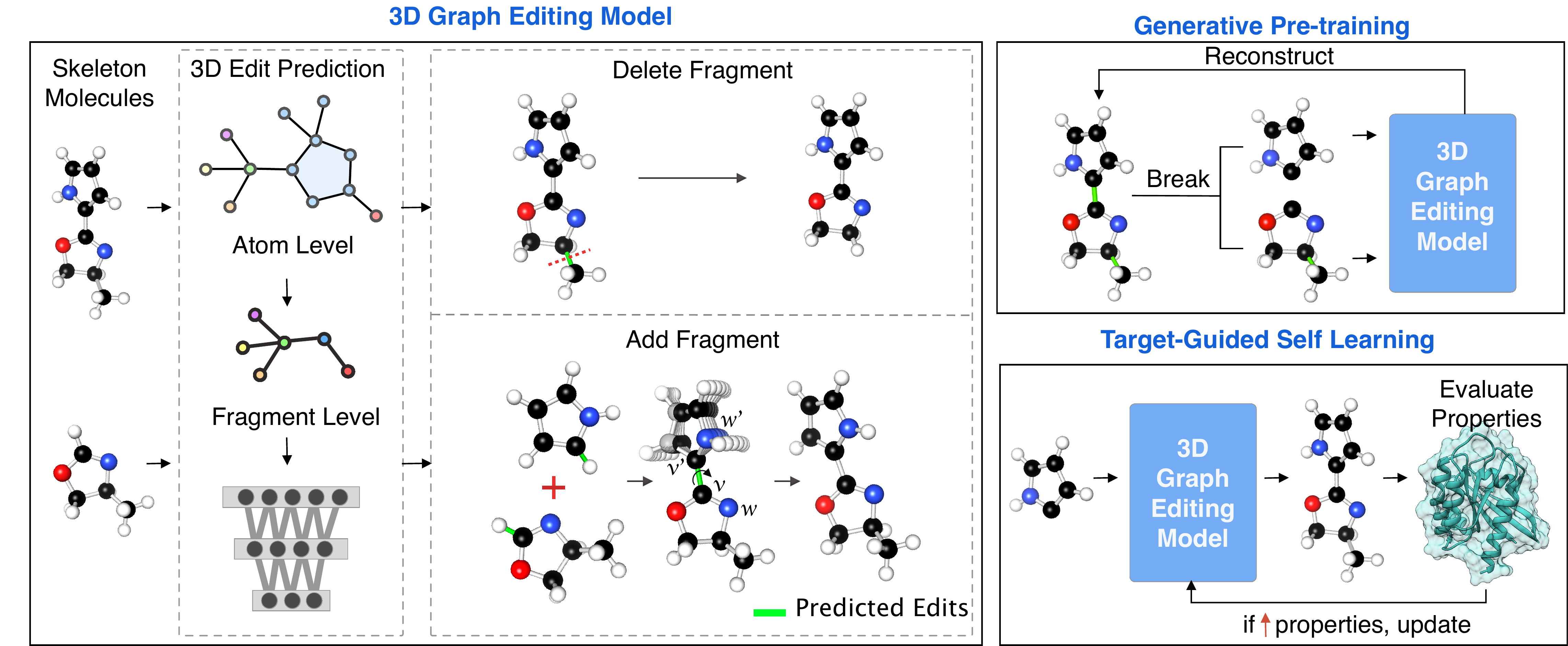}
    %\vspace{+0pt}
    \caption{Model Overview. \method contains three components. 3D graph editing model predicts the geometric edits which either add or delete a rigid fragment from the skeleton molecule. For \textit{add} operation, the skeleton molecule is linked with a rigid fragment using the predicted attaching sites and torsion angle (defined by four consecutive atoms, $w$, $v$, $v'$ and $w'$). For \textit{delete} operation, the predicted bond is broken.
    With the editing model, we use generative pre-training to reconstruct 3D ligands for learning target-independent properties. 
    The model is further finetuned using target-guided self-learning strategy, which use self-generated molecules with improved target-related properties to enhance target-awareness. }
    \label{fig:overview}
    \vspace{-2mm}
\end{figure}

SBDD task is formulated as: given a 3D target structure, generate diverse molecules satisfying desired properties. In this section, we address the procedure of \method for solving the SBDD task. 
As shown in Figure \ref{fig:overview}, \method contains three components: 3D graph editing model, generative pre-training, and target-guided self learning. 3D molecules are generated by adding or deleting 3D fragments with the graph editing model, which is first pre-trained with abundant 3D ligand molecules to capture target-independent properties, such as drug-likeness, synthesizability and stability, and then further fine-tuned using self-generated molecules with improved target-related properties to enhance its capability of generating high binding affinity ligands for a specific target.

\subsection{3D Graph Editing Model}\label{sec:edit}
3D graph editing model is developed to generate 3D molecules within the target binding site. In molecules, single bond rotation is the major cause of 3D conformation changes, therefore, we represent molecules as a group of linked rigid 3D fragments. We build a rigid fragment library by breaking non-terminal single bonds. The broken bonds are labeled as editable sites for molecular editing and a hydrogen atom is added to maintain the original valency. More details of the fragment library can be found in Appendix. %\ref{appendix:experiment_details}

\method places an initial seed molecule within a target binding site, and builds 3D molecules by adding or deleting rigid fragments iteratively as shown in Figure ~\ref{fig:overview} \textit{Left}. In order to achieve adding and deleting operations separately, the editable sites are categorized as addable sites, which correspond to the editable sites with hydrogen atoms attached, and deletable sites which correspond to those with non-hydrogen atoms attached. In adding operations, the model firstly selects an addable sites in the skeleton molecule (the molecule to edit). Then it chooses a 3D fragment from the fragment library, and an addable site in the fragment to attach to the skeleton molecule. Lastly, it decides the torsion angle between the two connected components. 
%To conduct this predicted edit, the hydrogen atoms in the selected addable bonds are removed resulting in two partial bonds, which are then connected and rotated to the predicted torsion angle. 
In deleting operations, the model chooses a deletable site in the skeleton molecule to break. A hydrogen atom is added to the broken bond to maintain the correct valency.

\paragraph{Parameterization of Editing Operations}
To achieve 3D molecular editing, we represent a 3D molecule using hierarchical graphs, which have a atom-layer and a fragment layer. A network $\phi$ is used to parameterize the hierarchical graphs. Let $x$ be the skeleton molecule, we select an edge in the fragment-level to edit as follows:
\begin{gather}
    \vs^\text{atom-node},\vs^\text{frag-node},\vs^\text{frag-edge} = \phi_1(x) \\
    \text{score}_{j, k}^{\text{skel}} = \textsc{MLP}_1(\vs_{j,k}^\text{frag-edge})\in \R \\
    p_\text{add}(r|x_\text{skel}) = \softmax(\{\text{score}_{j, k}^{\text{skel}}\}_{(j, k)\in E_a}) \\
    p_\text{delete}(r|x_\text{skel}) = \softmax(\{\text{score}_{j, k}^{\text{skel}}\}_{(j, k)\in E_d})
\end{gather}
where $\vs^\text{atom-node}$ is the skeleton molecule's atom-layer node hidden representations, $\vs^\text{frag-node}$ and $\vs^\text{frag-edge}$ are fragment-layer node and edge hidden representations respectively. ($j$, $k$) is a directed edge pointing from node $j$ to $k$. $E_a$ and $E_d$ are disjoint sets of editable edges in $x$ for addition and deletion respectively. MLP standards for multi-layer perceptrons. We sample a directed edge $r=(u_\text{skel}, v_\text{skel})$ from $\frac{1}{2}p_\text{add} + \frac{1}{2}p_\text{delete}$ as the predicted site to edit.

If the selected edge $r=(u_\text{ske}, v_\text{skel})$ is deletable, we replace the fragment on $u_\text{skel}$ side with hydrogen. Otherwise, $r$ is addable which means $u_\text{skel}$ is an hydrogen. We remove $u_\text{skel}$ and choose a fragment $f$ from the fragment library $H$ to add to $v_\text{skel}$:
\begin{gather}
    \text{score}_{f} = \textsc{MLP}_2(\vs^\text{frag-edge}_r)_{f}\in \mathbb{R} \\
    p_\text{frag}(f|x,r) = \softmax(\{\text{score}_{f}\}_{f \in H})
\end{gather}
We sample a fragment $f$ from $p_\text{frag}$. Then we need to select an edge $a$ in the fragment $f$ to attach to edge $r$ in the skeleton molecule $x$. The edge to attach in the fragment is determined jointly by $x$ and $f$:
\begin{gather}
    \vo^\text{atom-node},\vo^\text{frag-node},\vo^\text{frag-edge} = \phi_2(f) \\
    \bar{\vs}^\text{frag-node} = \textsc{MeanPool}(\vs^\text{frag-node}) \\
    \text{score}_{j, k}^\text{frag} = \textsc{MLP}_3(\textsc{Concat}(\bar{\vs}^\text{frag-node}, \vo^\text{frag-edge}_{j, k})) \in \R \\
    p_\text{attach}(a|x,r,f) = \softmax(\{\text{score}_{j, k}^\text{frag}\}_{(j, k)\in E_a^\text{frag}})
\end{gather}
where $E_a^\text{frag}$ is the set of addable bonds in fragment $f$. Then we sample an edge $a=(u_\text{frag}, v_\text{frag})$ to attach. Finally, we gather the atom-layer features of four atoms, $w_\text{skel}$, $v_\text{skel}$, $v_\text{frag}$, $w_\text{frag}$, around the new bond ($w_\text{skel}$ is the neighboring atom of $v_\text{skel}$ in the skeleton and $w_\text{frag}$ is that of $v_\text{frag}$ in the added fragment as shown in Figure \ref{fig:overview} \textit{Add Fragment})  and determine the torsion angle as a classification task which separates angles into discrete bins $\Gamma = \{0, 10, 20, \cdots 350\}$:
% \footnotetext{Note that $w_\text{skel}$ is the neighboring atom of $v_\text{skel}$ in the skeleton and $w_\text{frag}$ is that of $v_\text{frag}$ in the added fragment. $w_\text{skel}$, $v_\text{skel}$, $w_\text{frag}$, $v_\text{frag}$ is shown as $w$, $v$, $w'$, $v'$ in Figure \ref{fig:overview}}
\begin{gather}
    \vh^\text{angle} = \textsc{Concat}(\vs^\text{atom-node}_{w_\text{skel}}, \vs^\text{atom-node}_{v_\text{skel}}, \vo^\text{atom-node}_{w_\text{frag}}, \vo^\text{atom-node}_{v_\text{frag}}) \\
    \text{score}^\text{angle}_{\gamma} = \textsc{MLP}_4(\vh^\text{angle})_\gamma \in \mathbb{R}\\
    p_\text{angle}(\gamma|x,r,f,a) = \softmax(\{\text{score}^\text{angle}_\gamma\}_{\gamma \in \Gamma })
\end{gather}
% \footnotetext{$\Gamma = \{0, 10, 20, \cdots 350\}$.}
and sample an angle $\gamma$ from $p_\text{angle}$. We then connect the skeleton molecule and fragment molecule together by this torsion angle $\gamma$.
\paragraph{Parameterization of Hierarchical Graphs}
We develop a hierarchical message passing neural networks (HMPNNs) to parameterize the molecular graphs. HMPNNs contain two MPNNs as its atom-layer and fragment layer. A input molecule $m$ is represented as a graph $g=(\mA,\vq^\text{atom-node},\vq^\text{atom-edge})$ with $\mA$ as the adjacency matrix, $\vq^\text{atom-node}$ and $\vq^\text{atom-edge}$ as feature vectors of atoms and bonds. 
We pass the graph through the first MPNN to obtain atom-level representations:
\begin{equation}
    \vh^\text{atom-node}_u = \textsc{MPNN}_1(g)_u\in\mathbb{R}^d
\end{equation} 
where $\vh^\text{atom-node}_u$ is the atom-level hidden representation of node $u$. Note that atoms separated by rotatable single bonds are considered as belonging to different fragments. We then regard each fragment as a single node, which induces a fragment-level adjacency matrix $\mA'$, and obtain the fragment embedding $\vz^\text{frag-node}$ by aggregating features of atoms that belong to it using mean pooling. We preserve those edges between fragments with feature vectors $\vz^\text{frag-edge}$ initialized by fragment embeddings:
%\end{gather}
\begin{gather}
    \vz^\text{frag-node}_i = \textsc{MeanPool}_{u\in V_i}(\vh^\text{atom-node}_u) \in \mathbb{R}^d \\
    \vz^\text{frag-edge}_{j,k} = 
    \mW_1\cdot\textsc{Concat}(\vz^\text{frag-node}_j, \vz^\text{frag-node}_k) + \vk_1\in\mathbb{R}^d
    \intertext{where $V_i$ is the set of atoms in fragment $i$; $j$ and $k$ are adjacent fragments. The new graph $g'=(\mA',\vz^\text{frag-node},\vz^\text{frag-edge})$ is then passed to another MPNN to obtain fragment-level representations:}
    \vh^\text{frag-node}_i = \textsc{MPNN}_2(g')_i\in \mathbb{R}^d \\
    \vh^\text{frag-edge}_{j,k} = 
    \mW_2\cdot\textsc{Concat}(\vh^\text{frag-node}_j, \vh^\text{frag-node}_k) + \vk_2\in\mathbb{R}^d
    %\intertext{where $\vh^\text{frag-node}_{i}$ is the hidden representation of fragment $i$ and $\vh^\text{frag-edge}_{a,b}$ is the hidden representation of edge between fragment $a$ and $b$.}
\end{gather}
where $\vh^\text{frag-node}_{i}$ is the hidden representation of fragment $i$ and $\vh^\text{frag-edge}_{j,k}$ is the hidden representation of edge between fragment $j$ and $k$.

\subsection{Generative Pre-training for 3D Graph Editing Model}\label{sec:pretrain}
Due to the scarcity of target-ligand complexes, we propose to pre-train a generative model with 3D molecules to extract information of target-independent properties. 
% Since we are training a generative model without protein-ligand pairs, we want to learn much information from molecules themselves. 
% We are motivated by pre-training methods for text. 
% However, our problem setting is different because we have 3D molecules and 3D fragments. 
% When we connect a fragment to a molecule, the fragment can rotation relative to molecule's mounting point. 
We define a pre-training objective for 3D graph editing model using 3D molecular reconstruction. 
Notice that in this step, our pre-trained model does not contain information about a specific target. 
Therefore it will be able to generate valid molecules but may not bind tightly with a target. 

Given a drug-like 3D molecule $x$, we represent it as a fragment graph $g^\text{frag}$ with nodes as fragments defined in our fragment library and edges as bonds between fragments. The pre-training data of each molecule is then generated in an iterative manner. As shown in Figure \ref{fig:overview} \textit{Top Right}, at each step we randomly break an edge connecting a leaf node $f$ and the rest of the graph $g^\text{frag}_{\neg f}$, i.e., the deletable edges mentioned in Section \ref{sec:edit}. The remaining graph $g^\text{frag}_{\neg f}$ forms a new molecule $x_{\neg f}$ and we add $(x_{\neg f}, x)$ to the pre-training data. Then we repeat this operation on $x_{\neg f}$ until there is only a single fragment left.

Given a set of pre-training data $D_p=\{(x_{\neg f}, x)\}$, we train the 3D molecular editing model to maximize the likelihood of predicting the operation of adding $f$ on the corresponding edge of molecule $x_{\neg f}$, i.e., the reverse operation of deleting $f$ from the original molecule $x$,
\begin{align}
    \mathop{\arg\max}\limits_{\theta}~\frac{1}{|D_p|}\sum_{(x_{\neg f},x)\in D_p}\log p_\theta(x|x_{\neg f})
\end{align}
where $p_\theta$ is the model with parameter $\theta$. This pre-training stage enables the model to capture the relation between 3D molecular structure and general drug-like properties. 
\subsection{Target-Guided Bayesian Sampling with Self-Learning}\label{sec:sampling}
In this stage, we use the pre-trained 3D graph editing model to generate molecules for a given target. 
The overall idea is to use the 3D graph editing model in a Bayesian sampling framework (e.g. Markov chain Monte Carlo sampling). 
However, one issue is the pre-trained 3D graph editing model does not contain target-specific information, therefore it might not be able to generate molecules tailored for the target. 
To fix this issue, we use the samples generated during the procedure to further fine-tune the 3d graph editing model.

Given a pre-trained 3D molecular editing model and a target protein, we start with an initial molecule $x_0$ (e.g. methane CH4) and employ multi-chain annealed Bayesian sampling~\cite{kirkpatrick1983optimization} with a target-guided objective function to sample desired candidate ligands. For $i_{th}$ chain at step $t$, the model proposes an editing operation (adding or deleting a fragment) to modify $x_t^i$ to $x'$. The proposed $x'$ can either be accepted $x_{t+1}^i=x'$ or rejected $x_{t+1}^i=x_{t}^i$ as determined by an acceptance probability $\mathcal{A}(x',x_t^i) = \min(1, \exp(\frac{\mathcal{J}(x')-\mathcal{J}(x_t^i)}{T}))$ where $\mathcal{J}$ is the target-guided objective function and $T$ is the annealing temperature controlling how greedy the process is. Here we use a linear combination of three scores as the objective function:
\begin{align}
    \mathcal J(x) = \textsc{Vina}_\text{min}(x) + \alpha\log\textsc{QED}(x) + \beta\log\textsc{SAscore}(x),
\end{align}
where $\textsc{Vina}_\text{min}$ provides a target-aware score that measures the binding affinity between the target protein and the 3D ligand structure, while $\textsc{QED}$~\cite{bickerton2012quantifying} and $\textsc{SAscore}$~\cite{ertl2009estimation} are two target-independent scores that measure the drug-likeliness and synthetic accessibility of the candidate ligand respectively. Given a 3D target-ligand complex, we use AutoDock Vina \citep{Eberhardt2021} to conduct a quick local minimization and calculate the binding affinity as  $\textsc{Vina}_\text{min}$, which is much faster then the docking procedure used in AutoGrow and RGA. 
We discuss more details of Vina in Appendix. %\ref{appendix:vina}. 

During sampling, a dataset $D_t$ for target-guided self-training is collected on-the-fly. Denote $x$ and $x'$ to be the molecule before and after edit respectively. If the objective score of $x'$ is higher than $x$, molecule pair $(x,x')$ is added to the dataset $D_t$. We train our model simultaneously with the sampling using weighted maximum likelihood estimation (WMLE) as follows,
\begin{align}
    \mathop{\arg\max}\limits_{\theta}~\frac{1}{|D_t|}\sum_{(x,x')\in D_t}\lambda(x',x)\log p_\theta(x'|x)
\end{align}
where $p_\theta$ is the model and $\lambda(x',x)$ is a monotonic function indicating the score difference between $x'$ and $x$. Here we choose $\lambda(x',x) = \min\{ \mathcal{J}(x')-\mathcal{J}(x), 5\}$. WMLE injects more target information into the training signal than direct MLE.

\section{Results and Discussions} 
    \begin{table*}[t]
\small
    \caption{Performance comparison between structure-based drug design methods. 1000 molecules per target are generated by each method and the average and standard deviation values are reported. Top 1 results are highlighted in bold. \method achieves SOTA performance on Validity, Success Rate, High Affinity and median Vina score, while maintaining adequate Uniqness and Diversity.}
    \label{tab:main}
    % \resizebox{\columnwidth}{!}{%
    \begin{center}
    \resizebox{\textwidth}{!}{
    \begin{tabular}{ c | c | c  c c c c c c c}
    \toprule
    
    \multirow{2}{*}{Type} & {Method} 
    & Valid($\uparrow$) & Uniq ($\uparrow$) & Div ($\uparrow$) & High Aff ($\uparrow$) & Vina ($\downarrow$) & QED ($\uparrow$)& SA($\uparrow$) & Succ ($\uparrow$)\\
    & & (\%) & (\%) & & (\%) & (kcal/mol) & & & (\%)\\ 
    \midrule 
    \multirow{5}{*}{Cond.} 
    & liGAN & 96.3 $\pm$ 1.0 & 99.9 $\pm$ 0.1 & 0.889 $\pm$ 0.001 & 0.3 $\pm$ 0.3 & -5.91 $\pm$ 0.43 & 0.41 $\pm$ 0.17 &  0.59 $\pm$ 0.11 & 1.0 $\pm$ 1.1 \\
    & AR & 79.2 $\pm$ 19.5 & 44.5 $\pm$ 9.1 & 0.838 $\pm$ 0.040 & 51.8 $\pm$ 23.0 & -9.34 $\pm$ 1.47 & 0.54 $\pm$ 0.19 & 0.53 $\pm$ 0.18 & 16.5 $\pm$ 18.6 \\
    & GraphBP & 99.6 $\pm$ 0.1 & \textbf{100.0 $\pm$ 0.0} & \textbf{0.924 $\pm$ 0.001} & 8.2 $\pm$ 8.8 & -6.34 $\pm$ 0.97 & 0.41 $\pm$ 0.21 & 0.46 $\pm$ 0.15 & 1.0 $\pm$ 1.0\\ 
    & DESERT & \textbf{100.0 $\pm$ 0.0} & 99.9 $\pm$ 0.2 & 0.917 $\pm$ 0.007 & 45.2 $\pm$ 34.3 & -9.20 $\pm$ 1.17 & \textbf{0.64 $\pm$ 0.17} & 0.65 $\pm$ 0.13 & 47.3 $\pm$ 18.8\\
    & Pocket2Mol* & \textbf{100.0 $\pm$ 0.0} & \textbf{100.0 $\pm$ 0.0} & 0.902 $\pm$ 0.006 & 61.1 $\pm$ 23.1 & -9.77 $\pm$ 1.42 & \textbf{0.64 $\pm$ 0.14 }& 0.74 $\pm$ 0.11 & 68.2 $\pm$ 22.2\\
    \midrule 
    \multirow{3}{*}{Opt.} 
    & MARS & 99.8 $\pm$ 0.0 & 99.5 $\pm$ 0.3 & 0.915 $\pm$ 0.003 & 13.9 $\pm$ 20.0 & -7.63 $\pm$ 0.91 & 0.42 $\pm$ 0.23 & 0.75 $\pm$ 0.09 & 21.5 $\pm$ 11.1 \\
    & AutoGrow & \textbf{100.0 $\pm$ 0.0} & 99.7 $\pm$ 0.3 & 0.871 $\pm$ 0.025 & 11.8 $\pm$ 14.4 & -7.92 $\pm$ 0.61 & 0.34 $\pm$ 0.15 & 0.59 $\pm$ 0.07 & 13.2 $\pm$ 8.3 \\ 
    %& RGA & \textbf{100.0 $\pm$ 0.0} & \textbf{100.0 $\pm$ 0.0} & \textbf{0.936 $\pm$ 0.002} & 3.9 $\pm$ 3.4 & -6.07 $\pm$ 0.49 & 0.53 $\pm$ 0.16 & 0.70 $\pm$ 0.12 & 8.4 $\pm$ 4.3\\
    & RGA & \textbf{100.0 $\pm$ 0.0} & \textbf{100.0 $\pm$ 0.0} & 0.923 $\pm$ 0.004 & 8.8 $\pm$ 8.9 & -6.61 $\pm$ 0.52 & 0.49 $\pm$ 0.18 & 0.71 $\pm$ 0.11 & 16.1 $\pm$ 7.3\\
    & \method (L) & \textbf{100.0 $\pm$ 0.0} & 99.0 $\pm$ 2.1 & 0.885 $\pm$ 0.006 & 66.0 $\pm$ 15.7 & -10.00 $\pm$ 1.00 & 0.55 $\pm$ 0.19 & 0.77 $\pm$ 0.09 & 78.7 $\pm$ 14.6 \\
    & \method & \textbf{100.0 $\pm$ 0.0}  & 99.2 $\pm$ 1.3 & 0.880 $\pm$ 0.009 & \textbf{70.3 $\pm$ 14.2} & \textbf{-10.16 $\pm$ 1.00}  & 0.55 $\pm$ 0.19 & \textbf{0.78 $\pm$ 0.08} & \textbf{82.0 $\pm$ 13.1 }\\ 
    \bottomrule
    \end{tabular}}
    \end{center}
    \small{*Only 100 ligands per target are sampled for Pocket2mol due to its high computational cost. }
    \vspace{-2mm}
\end{table*}
    \subsection{Experiments}
\paragraph{Model Details} We use ChEMBL\citep{gaulton2017the}, a database of bioactive molecules, for rigid fragment library construction and pre-training. HMPNN model contains 6 atomic layers and 3 fragment layers. The atomic node features include atomic number, element type, charge and 3D coordinates, and the atomic edge features include bond type. The hidden layer node embedding has a size of 64. 
We experiment with two versions of \method, differed by the number of chains employed in Bayesian sampling. The light version, \method (L)  uses 1000 chains and the regular version \method uses 5000 chains. More model details and computational cost can be found in Appendix. %\ref{appendix:experiment_details}

\paragraph{Evaluation}
Following \citet{masuda2020ligan,pmlr-v162-liu22m,long2022DESERT}, we evalaute our method using 10 targets and all assessed models generate 1000 ligands for each target. We evaluate them using the following metrics:

\textbf{Validity (Valid)} denotes the percentage of molecules that are readable by RDKit and have all atoms in the same connected component.
\textbf{Uniqueness (Uniq)} is the percentage of unique molecules among all generated ones. 
\textbf{Diversity (Div)} measures the internal diversity of the generated molecules.

To evaluate the models' capability of generating bioactive molecules, we also use \textbf{High Affinity (High Aff)} and \textbf{Vina Score (Vina)} to quantify the binding affinity. High Affinity is defined as the percentage of generated molecules with higher affinity (lower Vina score) than the reference ligand. Unlike Vina minimization score used in the objective function, \textbf{Vina Score} is computed after a docking procedure for a more accurate binding energy estimation. For target-independent properties, we use \textbf{QED} and \textbf{SAscore (SA)} to quantify drug-likeness and synthesizability. As drug design requires the generated molecules satisfy multiple requirements simultaneously, we use \textbf{Success rate (Succ)} to evaluate the percentage of generated molecules that pass the predefined thresholds for the desired properties. We define a qualified molecule to have QED $\geq{0.25}$, SAscore $\geq{0.59}$, and Vina score $\leq{-8.18}$ kcal/mol. QED and SAscore thresholds are defined as the 10th percentile of approved drugs in DrugCentral \citep{ursu2019drugcentral} and the intuition is to cover majority of real drugs. The Vina score threshold corresponds to a binding affinity less than 1 $\mu$M, which is a widely used value to guarantee a moderate bioactivity in medicinal chemistry.

We compare \method with two types of SBDD models, namely \textbf{conditional generation (Cond.)} and \textbf{optimization-based methods (Opt)}. Conditional generation methods learn the 3D ligand distribution conditioning on target binding sites, which includes liGAN\citep{masuda2020ligan}, AR\citep{luo2021a}, GraphBP\citep{pmlr-v162-liu22m}, DESERT\citep{long2022DESERT} and Pocket2Mol\citep{peng2022pocket2mol}. On the other hand, optimization-based method uses target-ware objective to guide the optimization process. AutoGrow4\citep{spiegel2020autogrow4} and RGA\citep{fu2022reinforced} belong to this category and are designed for the SBDD scenario. Additionally, we convert MARS, an optimization algorithm for ligand-based drug design, to the SBDD settings by substituting ligand-based affinity predictor with Vina docking score. More detailed description of baselines can be found in Appendix.

\begin{figure}[t]
    \centering
    \includegraphics[width=\linewidth]{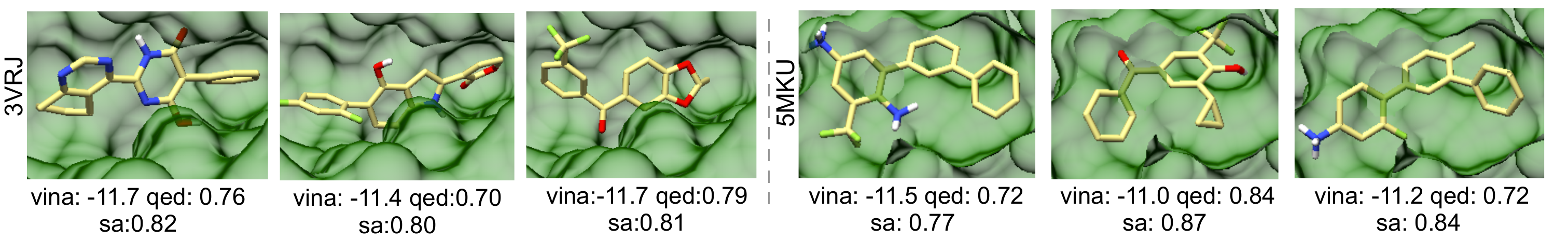}
    \vspace{-1mm}
    \caption{Target binding pose of \method generated molecules for 5MKU and 3VRJ proteins. The generated molecules demonstrate high Vina score, QED and SAscore.} 
    \label{fig:case}
    \vspace{-2mm}
\end{figure}

\begin{figure*}[t]
     \centering
     \includegraphics[width=\linewidth]{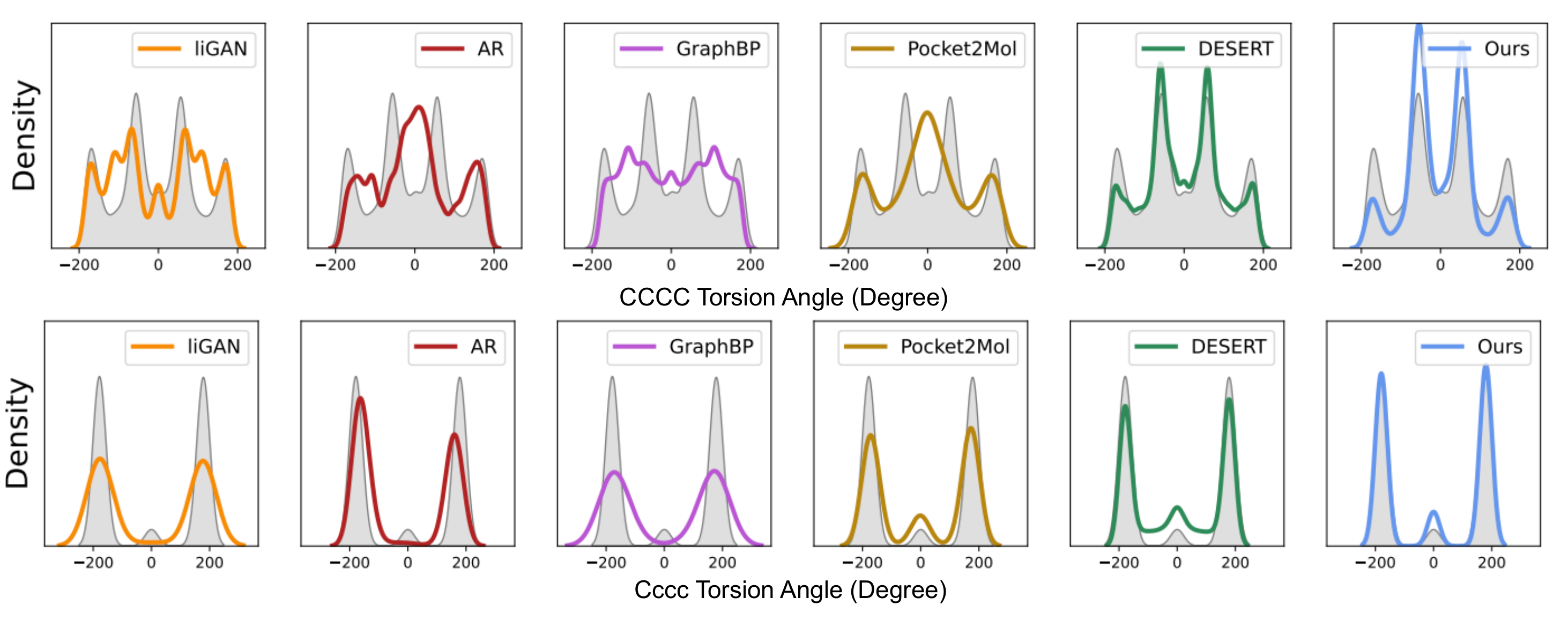}
    \caption{Angular distribution comparison for CCCC (upper panel) and Cccc (lower panel) torsion angles between CrossDocked2020 reference molecules and model generated molecules. DESERT and \method show better overlap with reference distribution. }
    \label{fig:torsion}
    \vspace{-4mm}
\end{figure*}

\subsection{Results and Analysis}

\paragraph{Main Result}
The performances of \method and baseline models are summarized in Table \ref{tab:main}, and the reported values are averaged over the 10 evaluated targets. Among baseline models, liGAN, AR and GraphBP generate atom types and positions using conditional variational autoencoder, autoregressive model and flow model. Since these models only generate atoms and require a post-processing algorithm to assign bonds, it is possible for them to generate incomplete molecules (the generate atoms cannot be connected into one molecule) or invalid substructures as demonstrated by the Valid metric in Table \ref{tab:main}. On the other hand, DESERT and \method utilizes fragments to build molecules, which can guarantee local validity. \method predicts the connectivity between fragments to generate complete molecules, and using pretraining and target-guided self-training to encourage validity. Similarly, DESERT uses pretraining and a greedy approach to link the generated fragments. Pocket2Mol generate bonds together with atoms, which also improves the molecular validity.  % More detailed performance on each target can be found in Appendix Table \ref{tab:set_a_ind}. 

\method achieves the best performance in generating high binding affinity molecules as reflected by High Affinity and Vina Score in Table \ref{tab:main}. liGAN, AR, GraphBP and Pocket2Mol use a supervised approach to learn the distribution of 3D ligands using  target-ligand complex in CrossDocked2020 \citep{francoeur2020three}. However, the data have mixed high affinity and low affinity molecules, which can influence the model performance. DESERT proposes to only use shape information of the target to guide the drug design. With the incomplete target information, the binding affinities of the generated molecules cannot be guaranteed. AutoGrow, RGA and MARS edit 2D molecules iteratively to optimize Vina docking score. Since their actions are defined in 2D space and may not correlate well with Vina score changes, the optimization performance is not satisfactory and their results are worse than some conditional generation models. 

Our method directly generates 3D molecules and adopts a target-guided self-training approach to accelerate the annealed sampling framework. As an optimization-based method, we demonstrate that defining action space in 3D is more efficient and effective for SBDD task.

 Additionally, \method achieves state-of-the-art performance on SAscore and Success Rate, and the latter simultaneously evaluate binding affinity, drug-likeness and synthesizability. Unlike binding affinity, drug-likeness and synthesizablity are target-independent and our method can simultaneously optimize all three metrics well. 

In Figure \ref{fig:case}, we show the docking poses of \method generated molecules for two targets, 5MKU and 3VRJ. The results indicate our generated molecules can form strong shape complementarity with the target binding site, which is the premise of high binding affinity.

\begin{figure}[t]
    \centering
    \includegraphics[width=\linewidth]{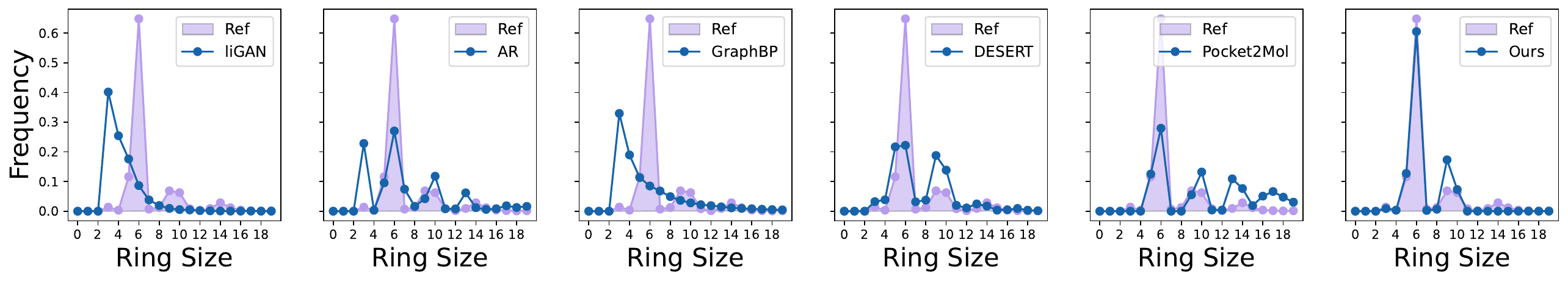}
    %\vspace{-1.9em}
    \caption{Frequency of different ring sizes for DrugCentral reference molecules and model generated molecules. \method has the best overlap with the reference frequency.} 
    \label{fig:ring}
    %\vspace{-1.9em}
\end{figure}

\paragraph{Additional Target-Independent Properties}
We analyze additional target-independent properties of the  generated molecules by 3D molecular generation methods here. To examine whether generated molecules have reasonable 2D structures, we compare generated molecules with those in DrugCentral Database \citep{ursu2019drugcentral}, which are either drugs or pharmaceuticals. We compare the frequency of different sized rings for molecules generated by SBDD models with the reference ones in Figure \ref{fig:ring}. liGAN, AR and GraphBP demonstrate high tendencies of generating 3-member rings. Meanwhile, DESERT and Pocket2Mol tend to generate larger ring systems. Among these models, \method shows the best alignment with reference drug molecules. 

In addition to 2D structures, we also evaluate the conformational stability of the generated molecules. Since bond length and angles are mostly encoded in the fragment vocabulary for fragment-based molecular generation models, such as DESERT and \method, we emphasize on the accuracy of torsion angles to make fair comparisons with atom-based models. We compare the distribution of two representative torsion angles, CCCC and Cccc, in Figure \ref{fig:torsion}. CCCC is a common torsion angle of aliphatic carbons (non aromatic), and Cccc is an angle around aromatic carbons. Here, we use the 3D ligands from the CrossDocked2020 database \citep{francoeur2020three} as reference. Normally, a torsion angle shows high frequency at a few preferred values, which corresponds to the energy-stable conformers for the rotatable bond of interest. As shown in Figure \ref{fig:torsion}, DESERT and \method achieve the best alignment with reference distribution. However, atom-based approaches, namely liGAN, AR, GraphBP and Pocket2Mol, are less satisfactory especially in the case of CCCC dihedral angle. In \method, we explicitly train the model to predict torsion angles between rigid fragments. DESERT uses fragments as building blocks for the molecular generation. Although torsion angles is not modeled explicitly in DESERT, it uses fragments containing rotatable torsion angles, which can benefit the generation of energy-stable conformers.

\begin{table}[t]
\small
    \centering 
    \caption{The influence of pre-training and target-guided self-training. }
    
    \label{tab:pretrain}
    \vspace{-2mm}
    \begin{tabular}{c c | c c c c c c c }
    \toprule 
    \multirow{2}{*}{Pretrain} & Self &
     Uniq ($\uparrow$) & Div ($\uparrow$) & High Aff ($\uparrow$) & Vina ($\downarrow$) & QED ($\uparrow$)& SA($\uparrow$) & Succ ($\uparrow$)\\
    & Learning & ($\%$) & & ($\%$) & (kcal/mol) & & & ($\%$)\\ 
    \midrule
     $\times$ & $\times$ & 100.0 & 0.895 & 10.7 & -8.83 & 0.68 & 0.67 & 59.5 \\ 
    $\times$& \checkmark & 100.0 & 0.895 & 16.3 & -9.07 & 0.68 & 0.68 & 67.0 \\ 
    \checkmark & $\times$  & 98.7 & 0.883 & 24.1 & -9.35 & 0.70 & 0.83 & 85.6\\ 
    \checkmark & \checkmark & 93.5 & 0.871 & 53.6 & -10.30 & 0.64 & 0.84 & 94.1 \\
    \bottomrule
    \end{tabular}
    \vspace{-4mm}
\end{table}

\subsection{Ablation Study}
In this section we study the contribution of pretraining and target-guided self-training using 5MKU target and \method (L) model as an example, and the results are summarized in Table \ref{tab:pretrain}. With pretraining, the Success Rate is improved from 59.5\% to 85.6\% on the without self-learning setting, and is improved from 67.0\% to 94.1\% on the with self-learning setting. Pretraining can significantly improve SAscore by 0.16 on both with and without self-learning settings, and also boost Vina Score. 
Meanwhile, target-guided self-learning has a major impact on binding affinities to the given target, reflected by Vina Score and High Affinity. With self-learning, Vina Score is improved on both with and without pretraining settings. 
In addition, High Affinity is increased by about 2 fold and 3 fold on with and without pretraining settings respectively. It should also be noted, pretraining can slightly decrease the Uniqueness and Diversity of the generated molecules, but the change is relatively small.

\section{Conclusion and Future Work} 
    In this paper, we propose \method, which adopts a sampling framework to generate 3D molecules in a target binding site and optimize desired properties. We propose a novel 3D graph editing model, and employs generative pre-training and target-guided self-learning to extract target-independent and -dependent properties respectively. \method achieves SOTA performance on Validity, binding affinity (High Affinity and Vina Score), SAscore and Success Rate, while maintaining adequate diversity, uniqueness and QED. In addition, \method generated molecules show strong agreement with reference molecules on both 2D and 3D molecular properties. Although \method has achieved state-of-the-art results on SBDD tasks, there is still space for improvement of this method and the evaluation of SBDD models in general. Practical drug design usually requires the designed molecules to form desired interactions, such as hydrogen bonding with a specific residue, which is a more sparse and harder objective for optimization. In addition, current evaluation metrics, such as QED and SAscore, has been questioned about their reliability, and better standards should be developed.

\bibliographystyle{plainnat}
\bibliography{ref}

\newpage
\begin{appendices}

   \section{Additional Related Works}
\label{appendix: related_work}
\textbf{Pre-training for Drug Design} Pre-training has been widely applied in 1D/2D drug design. \citet{Shi2020GraphAF} pre-train a flow-based autoregressive model with maximum likelihood estimation on ZINC dataset. \citet{krishnan2021novo} pre-train a generative model to encode the active site graph and grammar of small molecules. \citet{li2020learn} learn molecular representations from large-scale unlabeled molecules using an self-supervised strategy. Though popular in 1D/2D setting, pre-training is still under-explored in 3D structure-based drug design. 
\citet{luo2021a} and \citet{peng2022pocket2mol} pre-train atom-based and atom- and bond-based conditional generative models correspondingly following a mask prediction fashion using target-ligand complexes. 
DESERT~\cite{long2022DESERT} pre-trains a shape-to-molecule mapping on large-scale 3D molecular structures and achieves impressive results on zero-shot drug design. In comparison, our method is pre-trained for fragment-based molecular generation, and can generate molecules with higher binding affinities by incorporating a target-guided sampling with self-learning stage after pre-training.

\textbf{3D Conformation Prediction} 3D conformation prediction is defined as predicting the 3D pose of a molecule given its 2D structure, which specifies the atom types and their connectivity. Most of the deep models in this field learn to generate the atomic positions or the inter-atomic distance \citep{mansimov2019molecular, simm2020generative, xu2021learning, shi2021confgf}. These methods demonstrate initial success on relative small molecules, such as those in \texttt{GEOM-QM9} dataset \citep{axelrod2020geom}. However, their performances drop significantly when the molecular size increases, more specifically on drug-like molecules. Instead, \citet{gogineni2020torsionnet} and \citet{ganea2021geomol} propose to predict the rotatable torsion angles in order to generate 3D conformation. Modeling torsion angles shows a more stable performance when the number of rotatable bonds increases \citep{ganea2021geomol}. Unlike predicting distance metrics, which tends to overparameterize the degrees of freedom in the molecules, modeling torsion angles represents the molecular flexibility more straight forward and simplifies the exploration space of 3D conformers. In this work, we integrate this idea into our 3D geometric editing procedure and propose to generate new molecules by linking rigid fragments using flexible torsion angles.

\textbf{Reinforcement Learning (RL)} Reinforcement learning methods train agents to maximize the accumulated reward in an environment \citep{kaelbling1996reinforcement}. Policy gradient methods, a category of RL algorithms,  is the closest to our training procedure, estimate the gradients of agents' parameters with respect to sampled trajectories using policy gradient theorem \citep{policy_gradient,trpo,schulman2017proximal}. These methods have proven to be successful on many decision-making problems. While our self-training combined with annealed Bayesian sampling looks similar in the form to policy gradient methods, they are actually different. \method runs all chains simultaneously with an annealing temperature to control the acceptance rate of new molecules, while policy gradient methods usually sample trajectories in batches and accepts all transition edits. Overall, \method is simple to implement and empirically works well.
   \section{Autodock Vina Details}
\label{appendix:vina}
The bioactivity of a ligand with respect to a target protein largely depends on the tightness of the formed target-ligand complex, which can be measured by the minimum binding energy between two chemical compounds. As mentioned before, we use AutoDock Vina \citep{trott2010autodock}, one of the state-of-the-art molecular docking softwares, to estimate this energy.

Vina defines a scoring function to estimate the physical interactions in a target-ligand complex: 
\begin{equation} 
    \label{vina_dock}
    c = \sum_{i<j}f_{t_{i}t_{j}}(d_{ij})
\end{equation} 
where $i$ and $j$ are atom indices, $t_{i}$ and $t_{j}$ are the assigned atom types, $f_{t_{i}t_{j}}$ is the interaction function between atom types $t_{i}$ and $t_{j}$, and $d_{ij}$ is the interatomic distance between atom $i$ and $j$. The interaction function $f_{t_{i}t_{j}}$ is defined as a summation of steric interactions, hydrophobic interactions between hydrophobic atoms and hydrogen bonding between polar atoms. When atom $i$ and $j$ belong to both the ligand and the target, $c_{\text{inter} + \text{intra}}$ is the summation of intramolecular (within the ligand or the target) and intermolecular (between the ligand and the target) interactions. When atom $i$ belongs to the ligand and atom $j$ belongs to the target, $c_{\text{inter}}$ represents intermolecular interactions alone. 

Vina can predict the binding pose between a ligand and a target by searching for the global minimum of $c_{\text{inter} + \text{intra}}$. The process of predicting a target-ligand binding pose is called docking, and it is computationally expensive. If a ligand is placed in a relatively accurate position in the target binding site, then only a local minimization is needed to predict the binding pose. With a predicted target-ligand binding pose, Vina can estimate the binding energy $s$ following: 
\begin{equation}
    \label{vina_score}
    s = \frac{c_{\text{inter}}}{1 + wN_{\text{rot}}}
\end{equation}
where $N_{\text{rot}}$ is the number of active rotatable bonds in the ligand, and $w$ is the associated weight. 

The estimated binding energy by Vina can provide guidance for automatic drug design without the need of target-ligand complex data. However, 1D/2D deep models can only generate low dimensional molecules and requires the time-consuming docking module in Vina to predict the target-ligand binding pose in order to calculate the binding energy ($\text{Vina}_\text{dock}$), while \method can get rid of the global search of Vina, which makes it 100x faster. This is because 1) \method generates ligand in 3D space so that we can directly use local energy minimization ($\text{Vina}_\text{min}$) to estimate the binding energy of current pose and 2) $\text{Vina}_\text{min}$ informs the model of which 3D pose better fits the target.

   \section{Experiment Details}
\label{appendix:experiment_details}
\subsection{Model Details}
\paragraph{Rigid Fragment Library}
\begin{figure}[tb]
    \centering
    \includegraphics[width=\linewidth]{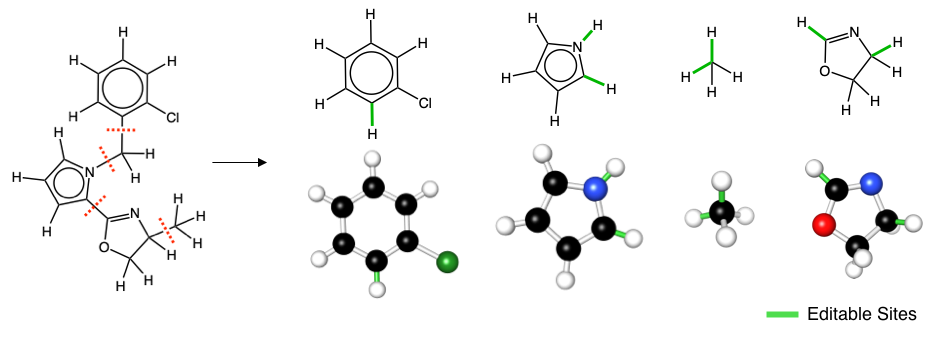}
    \caption{Mining 3D Rigid Molecular Fragments. 1) Non-terminal single bonds are broken to create 2D fragments and broken bonds are labeled as editable sites; 2) 3D conformations are generated for each fragment. }
    \label{fig:fragment}
\end{figure}
Prior approaches either represent a molecule as a string, a 2D graph, or a 3D graph with atoms as nodes. 
In this paper, we take a different representation. 
We represent a molecule using fragments as building blocks.
Larger molecules can be generated by adding fragments to partial molecules. 
Our goal is to first automatically discover proper fragments and then train a model to generate molecules by adding or deleting fragments. 

In our representation, each fragment is  a 3D graph whose nodes are atoms and edges are bonds. 
A molecule's 3D structure can be represented as a group of rigid 3D fragments linked by freely rotatable bonds. The bond rotation is quantified by its torsion angle defined using four consecutive atoms surrounding the bond of interest. Torsion angles are the major cause of conformational flexibility in 3D molecules. 

\method mines the basic fragments as follows. 
Single bond rotation is the major cause of torsion angle changes, as single bonds have lower rotational energy barrier comparing to other bond types. Therefore, we construct a rigid fragment library by breaking non-terminal single bonds in 2D molecules from the ChEMBL database \citep{gaulton2017the} as shown in Appendix Figure \ref{fig:fragment}. The broken bonds are labeled as editable sites for molecular editing and a hydrogen atom is added to maintain the original valency. The same 2D fragments with different editable sites are merged as one member in the library and the editable sites are combined. The 3D conformation of the resulted 2D fragments are generated using RDKit \citep{rdkit}. As fragments are designed to have limited rotational flexibility, each 2D fragment only corresponds to one or a few valid 3D conformers and each conformer is considered an independent rigid fragment. Using rigid fragments as building blocks can enhance the validity of the generated molecules and largely reduces the 3D conformational space.
The 1000 most frequent fragments in the ChEMBL database with no more than 10 heavy atoms are selected to construct the rigid fragment library. 

\paragraph{Pre-training Dataset}
We use a subset of 150K molecules from ChEMBL database to pre-train the 3D graph editing model. The original ChEMBL molecules are in 2D format and we use RDKit to generate its 3D conformers and optimize them to minimum energy. For each molecule, we randomly sample several iterative decomposition orders for pre-training. As a result, we obtain 1.2 million data points for pre-training. 

\paragraph{Training and Sampling}
HMPNN model contains 6 atomic layers and 3 fragment layers. The atomic node features include atomic number, element type, charge and 3D coordinates, and the atomic edge features include bond type. The hidden layer node embedding has a size of 64.
The editing model is pre-trained for 100 epochs. For target-guided sampling with self-training, multiple chains of annealed sampling proceed simultaneously and total 1000 sampling steps are performed. Temperature is annealed every 5 sampling step with rate 0.97. The maximum size of the dataset $D_t$ collected during sampling is 75K with first-in-first-out mechanism. We train the model every 5 sampling steps using $D_t$ with maximum 1000 gradient updates each time. For both pre-training and target-specific tuning, we use Adam optimizer with learning rate 3e-4 and set batch size to be 200. The generated molecules are restrict to contain no more than 40 heavy atoms. Finally, 1000 molecules are randomly selected from the last step.

\paragraph{Objective Function} We use Meeko \citep{meeko} to prepare ligand structures and MGLtools\citep{morris2009autodock4} to prepare protein structure preparation for AutoDock Vina\citep{Eberhardt2021} calculation. Exhaustivness=8 is used in Autodock Vina minimization. The negative of Vina minimization energy is used in our objective function. \textsc{QED}\citep{bickerton2012quantifying} and \textsc{SAscore} \citep{ertl2009estimation} are calculated using the implementations in RDKit \citep{rdkit}. We rescaled SAScore as 
 $(10 - \textsc{SAScore}) / 9$

\subsection{Evaluation}
\paragraph{Test Set}
We evaluate our method on drug design tasks using total 10 targets with PDB IDs of 1FKG, 2RD6, 3H7W, 3P0P, 3VRJ, 4CG9, 4OQ3, 4PS7, 5E19 and 5MKU, as in previous works~\citep{masuda2020ligan,pmlr-v162-liu22m,long2022DESERT}. The target 3D structures are downloaded from CrossDocked2020 \citep{francoeur2020three}. 

\paragraph{Baselines}
We compare \method with the following 3D drug design models and test their capability of generating 3D molecules within target binding sites:
\begin{itemize}[itemsep=1pt, leftmargin=10pt, parsep=0pt, topsep=1pt]
    \item \textbf{liGAN~\cite{masuda2020ligan}} represents molecules as atomic density grids, and uses conditional variational autoencoders in conjunction with a GAN loss to learn the 3D ligand distribution from existing ligand-target complexes. %The atom types and 3D coordinates are determined by optimizing the fitness to the generated atomic density grids.
    \item \textbf{AR~\cite{luo2021a}} develops a 3D drug design method, which models 3D atom occurrences within a target binding site and uses an auto-regressive approach to place atoms in the binding site. 
    \item \textbf{GraphBP~\cite{pmlr-v162-liu22m}} encodes the context (protein and current ligand) and places new atoms sequentially based on the contextual features while preserving the equivariance property. 
    \item \textbf{DESERT~\cite{long2022DESERT}} is a zero-shot 3D drug design method that is pre-trained on a large-scale unlabeled dataset and designs candidate ligands through a sketching-generating procedure.
    \item \textbf{Pocket2Mol~\cite{peng2022pocket2mol}} uses an equivariant generative network to model the 3D protein pockets and to sample ligands efficiently conditioned on pocket features.
    \item \textbf{AutoGrow4}~\cite{spiegel2020autogrow4} uses genetic algorithm to optimize 2D molecules, and employ molecular docking score as its objetive.
    \item \textbf{RGA} ~\cite{fu2022reinforced} is developed based on AutoGrow4 and uses reinforcement learning to improve the optimization efficiency of genetice algorithm. 
    \item \textbf{MARS} ~\cite{xie2021markov} uses Markov chain Monte Carlo sampling to generate 2D molecules and employ an adaptive training procedure to accelerate the sampling efficiency. 
\end{itemize} 

\paragraph{Computational Cost}
Conditional generative model only needs to train once, and can be generalized for different targets. The training time of Pocket2Mol is 72 hrs and AR is 62.5 hrs using a Nvidia v100 GPU. On ther other hand, optimization-based methods need to be trained for each target. We summarize the average computational cost of optimization in Table \ref{tab:opt_run_time}. 
Since \method directly generate 3D molecules, Vina minimization algorithm is used. However, other baselines generate 2D molecules instead, and require Vina docking oracle to predict the target-ligand binding pose alone binding affinity estimation. For MARS, we use 1000 chains and sample 100 steps. For AutoGrow, we run 100 generations with 1000 molecules to seed the next generation, 1000 mutants and 1000 crossovers per generation. 
For RGA, we run 10 generations with 20 molecules to seed the next generation, 500 crossovers and 500 mutants per generation. 
It should be noted, AutoGrow is the only model in Table \ref{tab:opt_run_time} which does not use deep learning to guide the optimization and does not spend time on model training. 
According to this table, \method is able to finish more oracle calls per unit time, which provides it with more intense target-aware signals during training.

\begin{table}[th]
\centering 
   \caption{Computational Cost of Optimization-Based Models}
   \label{tab:opt_run_time}
   \resizebox{\linewidth}{!}{
   \begin{tabular}{c|c c c c c}
   \toprule 
   Model & Run Time (hrs) & Oracle Type & $\#$Oracle Calls & Time/Orcale (s) & $\#$CPUs \\
   \midrule
   MARS & 55.5 & Vina Dock & 1 $\times$ $10^5$ & 1.998 & 64 \\
   AutoGrow & 11.88 & Vina Dock & 3 $\times$ $10^5$ & 0.143 & 64 \\ 
   RGA & 15.95 & Vina Dock & 1 $\times$ $10^4$ & 5.629 & 96 \\ 
   \method (L) & 38.06  & Vina Minimization & 1 $\times$ $10^6$ & 0.137 & 64 \\ 
   \method & 102.60 & Vina Minimization & 5 $\times$ $10^6$ & 0.074 & 64 \\
   \bottomrule
   \end{tabular}
}
\end{table}
\end{appendices}

\end{document}